\documentclass[aps,prb,twocolumn,superscriptaddress,showpacs,floatfix]{revtex4-2}

\usepackage{amsmath,amssymb,amsfonts,float,graphics,epsfig,epstopdf,color,verbatim,tabularx,bm,multirow,appendix}
\usepackage{CJKutf8}
\usepackage[utf8]{inputenc}
\usepackage[T1]{fontenc}
\usepackage{xcolor}
\usepackage{dsfont}
\usepackage{textcomp}
\usepackage{yfonts}
\usepackage{bm}
\usepackage{subfigure}
\usepackage{mathrsfs}
\usepackage{graphicx}
\usepackage{verbatim}
\usepackage{hyperref}
\usepackage{multirow}
\usepackage{braket}
\usepackage[normalem]{ulem}
\usepackage{tikz}
\usetikzlibrary{calc}
\usetikzlibrary{shapes.multipart}
\usepackage[commandnameprefix=ifneeded,defaultcolor=red]{changes}

\graphicspath{ {images/} }

\newcommand{\comments}[1]{}

\newcommand{\stkout}[1]{\ifmmode\text{\sout{\ensuremath{#1}}}\else\sout{#1}\fi}


\begin{document}

\begin{CJK*}{UTF8}{gbsn}
\title{An integrated theoretical and numerical approach to understand modern experiments on quantum magnetism}

\author{Zi Yang Meng \CJKfamily{bsmi}(孟子楊)}
\email{zymeng@hku.hk}
\affiliation{Department of Physics, The University of Hong Kong, Pokfulam Road, Hong Kong}
\affiliation{State Key Laboratory of Optical Quantum Materials, The University of Hong Kong, Pokfulam Road,  Hong Kong}

\author{Cristian D. Batista}
\email{cbatist2@utk.edu}
\affiliation{Department of Physics and Astronomy, University of Tennessee, Knoxville, TN 37996, USA}
\affiliation{Neutron Scattering Division, Oak Ridge National Laboratory, Oak Ridge, TN 37831, USA}

\author{Shiliang Li (李世亮)}
\email{slli@iphy.ac.cn}
\affiliation{Beijing National Laboratory for Condensed Matter Physics, Institute of Physics, Chinese Academy of Sciences, Beijing 100190, China}
\affiliation{School of Physical Sciences, University of Chinese Academy of Sciences, Beijing 100190, China}

\begin{abstract}
In recent decades, the study of quantum magnets, which feature unconventional behaviour such as exotic quantum phase transitions and quantum spin liquids, and unconventional magnetic states of matter, has made remarkable progress. However, each of the three foundational pillars --- numerical simulations, analytical methods, and, to a lesser extent, materials synthesis and experiments --- often tends to view itself as the primary driver of the field. Even through the need for collaboration among theory, numerics and experiment to understand the complex phases of quantum magnets is well established, yet, in our view there remains a persistent perception from experts in one area that the other two serve merely as supporting tool, primarily useful for validating the dominant ideas of one specialty, and less relevant to shaping the underlying scientific narrative. We refer to this mindset as the "pride and prejudice" in modeling and understanding modern quantum magnetism. In this article, we advocate for a different, more integrated approach to overcome the challenges faced by quantum magnetism researchers. We argue that this alternative mindset has already started to advance the understanding of several important quantum magnetic models and their materials realizations. 
\end{abstract}

\date{\today}
\maketitle

\noindent{\textbf{\large Introduction}}\\

Quantum magnets have long been a focus for condensed matter research. Theoretical models of spins on a lattice have provided many important physical insights, including predictions of non-trivial quantum effects such as topological order. Substantial experimental efforts have subsequently been dedicated to identify materials that exhibit such phenomena. These efforts have led to advances on multiple fronts, from the development of new theoretical frameworks to substantial progress in numerical algorithms, simulation techniques, materials synthesis, and experimental probes. 

Over time, the sophistication and complexity of analytical, numerical and experimental methods has continued to grow. Although in the past each of these approaches to studying quantum magnetism could be pursued independently, substantial advances in understanding now increasingly depend on a combination of experimental and theoretical insights. 

This article aims to highlight the growing need for the field of quantum magnetism to pursue an integrated research approach -- one that strengthens the synergy among theory, numerical simulation, and experiment -- to transcend the "pride and prejudice"~\cite{austenPride1918} mindset sometimes found in the community, and to advocate for a broader appreciation of how such cross-disciplinary collaboration accelerates discovery.

The value of this integrated approach is especially evident in the collective effort to identify and understand exotic phases, including evidence of quantum spin liquids, where quantum fluctuations destroy magnetic order. These phases are expected to exhibit fractionalization, whereby new quasiparticles emerge that carry parts of the fundamental electronic degrees of freedom, such as separate particles carrying only charge or spin, or even a fraction of them. These fractionalized quasiparticles interact among themselves via emergent gauge fields.
A major challenge in the search for spin liquids lies in the absence of definitive experimental `smoking gun' evidence. There have been many efforts to identify clear signatures for fractionalization but very often subsequent research found that they can also be mimicked by other less exotic effects such as disorder. Recent progress has instead relied on the indirect identification of spin-liquid behaviour through quantitative predictions of dynamical properties measured using spectroscopic probes, such as electron spin resonance, scanning tunnelling microscopy and inelastic neutron scattering. 

\section*{Success in one dimension}
The study of quasi-one-dimensional magnets has been an important success story for quantum magnetism, and one where independent theoretical and experimental efforts could make substantial progress. Early field theoretical work predicted that the constraints of one dimensional motion mean that the concept of interacting single particles breaks down and is replaced with a so-called Luttinger liquid with fractionalized excitations~\cite{haldane1981}.

The Luttinger liquid concept was validated by comparing theoretical predictions for the dynamical spin structure factor with inelastic neutron scattering data on long spin chains~\cite{lakeQuantum2005,coldeaQuantum2010}. A combination of numerical methods with analytical theoretical insights about integrable systems enabled an accurate reproduction of the scattering continuum, including its energy and momentum dependence under different applied magnetic fields. 

It is instructive to examine the key factors that enabled the successful discovery and characterization of magnetic Luttinger liquids. The first crucial element was the ability to synthesize quasi-one-dimensional spin systems with an exceptionally low degree of disorder, providing clean experimental platforms where intrinsic quantum effects are not masked by extrinsic factors \cite{Lake2013,Mourigal2013}. 

Secondly, accurate spin Hamiltonians for theoretical modelling could be constructed because many spin-$1/2$ chain materials are well described by a nearest-neighbour antiferromagnetic Heisenberg model. This meant that the appropriate model parameters could be easily determined by comparing experimental observation in inelastic neutron scattering measurements and model calculations with ab-initio methods as well as field theories.

A third important factor was the availability of powerful theoretical tools for computing dynamical spin correlation functions on long chains. These included both analytical methods such as the Bethe ansatz, and numerical techniques such as quantum Monte Carlo (QMC)~\cite{shaoNearly2017}, as well as time-evolution extensions of the density matrix renormalization group (DMRG)~\cite{haegemanUnifying2016}. By minimizing both the uncertainty in the underlying spin Hamiltonian and the errors in approximating its dynamics, it became possible to unambiguously reveal the fractionalized nature of the elementary excitations.

It is natural to ask why the same approach has not achieved the same success in the search for spin liquids in two- and three-dimensional materials. Some challenges already emerge in the one-dimensional case~\cite{SavaryL17,Knolle2019}. Even the well-understood material realizations of Luttinger liquids are quasi-one-dimensional magnets that  develop long-range magnetic order at sufficiently low temperatures~\cite{Lake2013,Mourigal2013}. As a result, the low-energy part of the inelastic neutron scattering spectrum features more conventional magnon excitations rather than the continuum of fractionalized excitations predicted by idealized field-theoretic models. 

In higher-dimensional materials, all three key ingredients that enabled progress in one dimension become much more challenging. First, the synthesis of low-disorder materials is more difficult, and two- and three-dimensional quantum spin liquids appear to be more sensitive to disorder. This makes it harder to realize clean experimental platforms where intrinsic quantum spin liquid behaviour can be isolated. 

Second, model extraction becomes less reliable, as many spin-liquid candidates in higher dimensions are governed by complex Hamiltonians with multiple competing interactions. Even with the most advanced \emph{ab initio} methods~\cite{villanovaFirstprinciples2023,LiuJ22,HeringM22}, the resulting parameter estimates often carry substantial uncertainties.

Third, the analytical and numerical methods used to probe dynamical properties are generally uncontrolled in the absence of small parameters, or they suffer from systematic limitations. Field-theoretical approaches offer valuable qualitative insights into the long-wavelength behaviour of magnetic systems, but often rely on idealized or oversimplified models. By contrast, the determination of realistic lattice Hamiltonians requires a close integration of theory with experimental data, complemented when possible by advanced \emph{ab initio} analysis. 

Even with a lattice Hamiltonian to hand, numerical schemes face challenges. For example, tensor network methods suffer from an exponential growth of entanglement, and, except in special cases, QMC simulations are afflicted by the minus sign problem, in which samples with negative weight make it difficult to perform a convergent calculation.

Faced with such difficulties, we believe that progress in quantum magnetism requires much closer collaboration between disciplines. As the following examples demonstrate, intensive combined common effort involving analytical theory, numerics, and experiment has led to substantial experimental and theoretical advances in several important quantum magnetic models. We argue that further integration and collaboration is not only essential for quantum magnetism, but also pivotal for advancing the broader field of quantum many-body physics.\\



\section*{\textbf{
Triangular lattice Quantum Ising magnets}} 

\begin{figure*}[htp!]
\includegraphics[width=\textwidth]{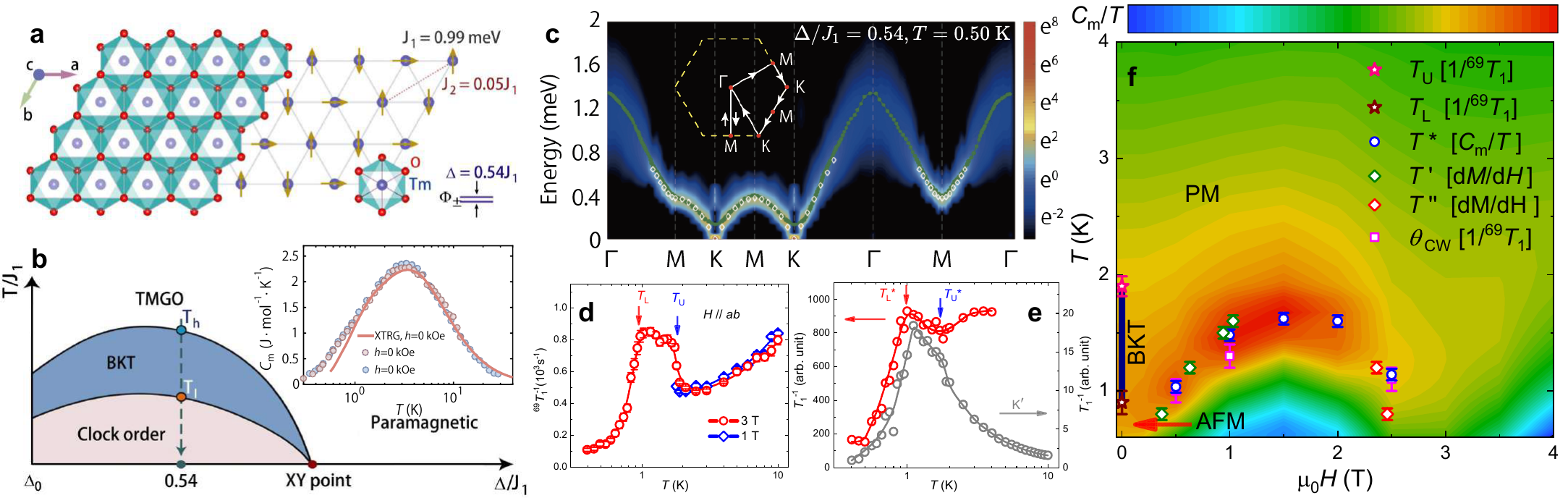}
\caption{\textbf{Triangular lattice quantum Ising magnet TmMgGaO$_4$.} (a) The Tm$^{3+}$ ions have an energy splitting $\Delta$ between two lowest non-Kramers electronic energy levels, labeled $\Phi_{\pm}$. These two levels constitute an effective spin-1/2 triangular lattice Ising model on the triangular lattice, with nearest neighbour and next nearest neighbour coupling parameters $J_1$ and $J_2$, respectively. An illustration of spin structure in the low temperature clock phase of TmMgGaO$_4$ (TMGO) is provided, where the spin-up and spin-down arrows are along the out-of-plane magnetic easy c-axis and the horizontal arrow illustrates spins in a superposition of spin up and down down states. (b) A schematic phase diagram of quantum triangular lattice Ising model, as a function of temperature $T$ and $\Delta$ normalised by $J_1$. The quantum critical point on the horizontal axis (red dot) has emergent spin XY symmetry, which extends into a Berezinskii-Kosterlitz-Thouless (BKT) phase at finite $T$ (blue region). The vertical arrowed line along $\Delta/J_1=0.54$ represents the relevant parameter range for TMGO, with two BKT transitions at $T_L$ (red dot) and $T_U$ (blue dot). The inset shows specific heat ($C_\text{m}$) measurements and their comparison with the thermal tensor renormalization group (TRG) computations for the triangular lattice Ising model. This comparison was used to determine the optimal set of the model parameters for quantum Monte Carlo (QMC) calculations~\cite{liKosterlitz2020}. (c) QMC calculated spin spectrum at $T=0.5$ K~\cite{liKosterlitz2020}. The computed spectrum agrees well with the experimental inelastic neutron scattering results taken from Ref.~\cite{shenIntertwined2019}, whose peak positions are shown as the green dotted line. (d) NMR measurements of the ${}^{69}$Ga spin relaxation rate 1/${}^{69}T_1$ vs. temperature ($T$) measured under in-plane magnetic fields $H$ of 3 T and 1 T. A plateau feature, characterizing the anticipated BKT phase, is observed between $T_L \approx 0.9$ K and $T_U \approx 1.9$ K~\cite{huEvidence2020}. (e) QMC computed $1/T_1$ data from the dynamical spin–spin correlation function with contributions from all momentum points in the Brillouin zone (left scale) and from momenta in the vicinity of the K' Brillouin zone point (right scale)~\cite{huEvidence2020}. (f) Experimental phase diagram of the material under an applied magnetic field.  The solid vertical line illustrates the BKT phase between $T_U$ and $T_L$, while the red arrow indicates the antiferromagnetic (AFM) clock order regime~\cite{liaoPhase2021}. The temperatures corresponding to the maximum of the $C_\text{m}/T$ ($T^{\ast}$), maximum of $dM/dH$ ($T^{\prime\prime}$) where $M$ is the material magnetization, and the Curie–Weiss temperature $\theta_{CW}$ all collapse to the same phase boundary between the BKT-like regime and the AFM phase~\cite{huEvidence2020}.} 
\label{fig:figTMGO}
\end{figure*}


The triangular lattice Quantum Ising (TLI) material TmMgGaO4 (TMGO), 
is, due to its Ising character, a rare case in two dimensions where unbiased, large-scale QMC and thermal tensor renormalization group (TRG) can be performed, which enabled very close connections between numerical and specific heat and inelastic neutron scattering measurements~\cite{liKosterlitz2020}. 

As shown in Fig.~\ref{fig:figTMGO}(a), \emph{ab initio} analysis of the experimental data indicates that TMGO is an Ising-type triangular antiferromagnet~\cite{cevalloAnisotropic2018}. Each Tm$^{3+}$ ion carries a total spin-orbit moment of 6, but  the crystal electric field selects a non-Kramers ground-state doublet as the effective magnetic degree of freedom.
Because the crystal field gap between second and third levels is about 400 K~\cite{liPartial2020}, the magnetic ion can be regarded as an effective spin-1/2 moment. The local trigonal crystal field induces an energy splitting $\Delta$ within the doublet, which acts as an intrinsic transverse field on the effective spin 1/2~\cite{cevalloAnisotropic2018}. 

Therefore, Tm$^{3+}$ ions contribute highly anisotropic Ising-type moments with $J_z = \pm 6$ along the $c$-axis out of the two-dimensional plane of the triangular lattice, with an effective out-of-plane $g$-factor $ g^{zz} \sim 13.2$~\cite{liKosterlitz2020}. 
The material is thence described by  the TLI model
\begin{equation}
H_{TLI}=J_1\sum_{\langle i,j \rangle} S^{z}_iS^{z}_j + J_2 \!\!\! \sum_{\langle\langle i,j \rangle\rangle} \!\!\!  S^{z}_iS^{z}_j -\sum_i(\Delta S^{x}_i + h  S^z_i),
\label{eq:eq1}
\end{equation}
where $S^{x,y,z}_i$ are the spin operators at site $i$, the nearest-neighbour (next-nearest-neighbour) coupling is $J_1$ ($J_2$), $\Delta$ is the energy splitting the two lowest non-Kramers $\Phi_\pm$ levels, and $h= H^z g^{zz} \mu_B $, with  $H$ being the $z$-component of the external magnetic field and $\mu_B$ is the Bohr magneton. 

The generic phase diagram of the TLI model (Fig.~\ref{fig:figTMGO}~(b)) is well-established in the field theoretical and numerical literature~\cite{isakovInterplay2003,liaoPhase2021}. It includes a clock-ordered phase (see Fig.~\ref{fig:figTMGO}(a)) that breaks the $C_6$ symmetry of the triangular lattice for transverse field values below a critical threshold $\Delta < \Delta_c$ and below a Berezinskii-Kosterlitz-Thouless (BKT) transition temperature $T_L$. Above the clock-ordered phase,  in the temperature range $T_L <T <T_U$, there exists a critical region characterized by power-law spin-spin correlations and an emergent XY symmetry. 

To connect the generic physics of the model to the specific material TMGO, it is crucial to determine the corresponding TLI model parameters, particularly the ratio $\Delta/J_1$. The availability of large-scale QMC and TRG calculations meant that numerical simulations of the specific heat and neutron-scattering spectrum could be quantitatively compared with experimental data (Fig.~\ref{fig:figTMGO}(b) inset and (c)). These comparisons were then used to precisely determined of the model parameters of TMGO~\cite{liKosterlitz2020}. 

The extracted parameters placed TMGO in a region of the phase diagram with accessible clock and BKT phases, with $\Delta/J_1=0.54$.  In contrast, simple mean-field calculations of the model estimated 
$\Delta/J_1 = 1.36$~\cite{shenIntertwined2019} and would place the material in the disordered paramagnetic phase as $\Delta > \Delta_c \sim 0.8J_1$, the location of the quantum critical point. This suggests that the analytic mean-field treatment fails to capture quantum fluctuations intrinsic to the quantum TLI model and the material itself. The unbiased quantum many-body TRG+QMC scheme is necessary to explain the observed inelastic neutron scattering spectrum.

Furthermore, the QMC simulations of the TLI model in Eq.~\eqref{eq:eq1}, for Hamiltonian parameters determined in Ref.~\cite{liKosterlitz2020}, quantitatively predicted the existence of a BKT phase between the high temperature disordered phase and a low temperature antiferromagnetic phase.
This predicted phase was later measured and confirmed using nuclear magnetic resonance (NMR)~\cite{huEvidence2020} .

 Comparisons between the observed NMR signal (Fig.~\ref{fig:figTMGO} (d)) and predictions using QMC (Fig.~\ref{fig:figTMGO} (e)) demonstrate that the numerical data match NMR experiments.

\begin{figure*}[htp!]
\includegraphics[width=\textwidth]{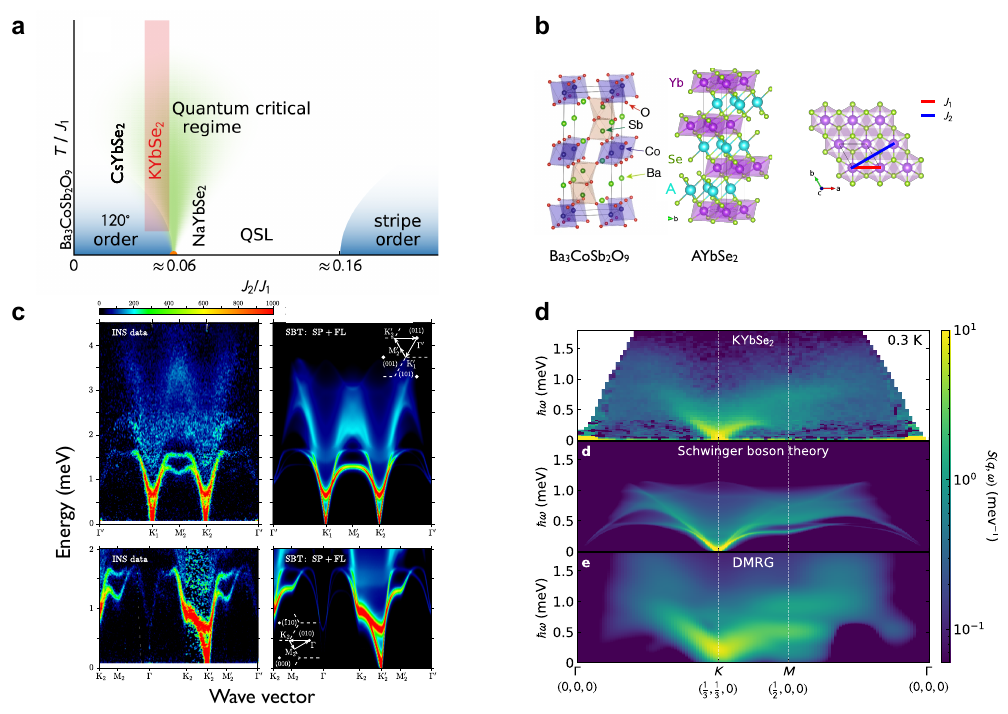}
\caption[Triangular Lattice Heisenberg Antiferromagnets.]{\textbf{Triangular Lattice Heisenberg Antiferromagnets.} (a) Schematic phase diagram of the $J_1$-$J_2$ Heisenberg model based on numerical results~\cite{Kaneko14,Zhu15,Hu15,Iqbal16,Saadatmand16,Zhu18} featuring a large quantum spin liquid (QSL) phase between two magnetically ordered phases (blue regions). The parameter regimes corresponding to several candidate materials are illustrated.  
(b) Structure of the transition metal oxide Ba$_3$CoSb$_2$O$_9$ and the Yb delafossite AYbSe$_2$ with A being an alkali metal.  (c) Comparison between inelastic neutron scattering measurements of Ba$_3$CoSb$_2$O$_9$~\cite{Macdougal20} (left) and the Schwinger boson theory calculation of the scattering cross section~\cite{Ghioldi22} (right). The low-energy, sharply defined magnon modes are interpreted as two spinon-bound states, while the continuum scattering is dominated by the two-spinon continuum. (d) Comparison of inelastic neutron scattering data measured in AYbSe$_2$ with the dynamical spin structure factor obtained using Schwinger boson theory and density matrix renormalization group (DMRG) calculations, reproduced from Ref.~\cite{scheieProximate2024}. } 
\label{fig:fig1}
\end{figure*}



This TLI case study underscores the power of combining experimental input with controlled many-body methods to refine and validate microscopic models. Theoretical insights into the BKT phase, together with numerical methods for comparison with specific heat and inelastic neutron scattering provide a coherent framework for interpreting the data. The experiment-driven numerical methods then produced a prediction for the magnetic field range of the finite temperature BKT phase, leading to its observation using NMR. These elements converge to yield the complete phase diagram shown in Fig.~\ref{fig:figTMGO}(f).\\

\section*{\noindent{\textbf{Triangle lattice Heisenberg antiferromagnets}}}

A longer-standing and more fundamental problem in quantum magnetism that highlights the need for a balanced and collaborative approach 
is the triangular lattice Heisenberg antiferromagnet (TLHAF). The geometric frustration of the lattice led to the proposal of  the resonating valence bond (RVB) spin liquid state state~\cite{Anderson73} over 50 years ago. 

Although the nature of the TLHAF ground state remained a topic of debate for many years~\cite{Zheng06}, a sequence of numerical works~\cite{Huse88,Bernu92,Capriotti99} have
provided evidence in favour of 120$^{\circ}$ long range 
magnetic order (LRMO) with a relatively small
ordered moment (41\% of the full moment)~\cite{Capriotti99}. The reduction of the ordered moment indicates strong quantum fluctuations, which, in a semi-classical treatment, are reflected by large corrections to the magnetic excitation spectrum, including extensive decay of magnon excitations across much of the Brillouin zone~\cite{Starykh06,Zhitomirsky13}.

In line with the proposal of a non-classical RVB state, early studies of the low-temperature properties of TLHAF based on effective quantum field theories suggested the need for alternative descriptions beyond the semiclassical approach~\cite{Read91,Chubukov94}. This was further supported by numerical work~\cite{Kaneko14,Zhu15,Hu15,Iqbal16,Saadatmand16,Zhu18} on an extended model that includes nearest-neighbour ($J_1$) and next-nearest-neighbour ($J_2$) exchange interactions. These results indicate that a nonzero $J_2$ leads to a continuous quantum phase transition into a spin liquid state at the quantum critical point $J_2/J_1 \approx 0.06$,  as depicted in the phase diagram in  Fig.~\ref{fig:fig1}~(a). 

These studies prompted a series of inelastic neutron scattering experiments~\cite{Ma16,Ito17,Kamiya18,Macdougal20} on the quasi-two-dimensional nearest-neighbour TLHAF Ba$_3$CoSb$_2$O$_9$, shown in Fig.~\ref{fig:fig1}~(b), which revealed an anomalous scattering continuum and a strong renormalization of the single-magnon dispersion, confirming the expected breakdown of the semiclassical expansion due to proximity to the quantum critical point. In this regime, magnons should be described as bound states of two fractionalized spinons, which are the relevant quasiparticles in a spin liquid state. At higher energies, the scattering continuum continuously evolves into the two-spinon continuum expected in a spin liquid phase.

Large-$N$ parton approaches for studying spin liquid phases, where $N$ denotes the number of parton flavors, are well-suited to  account for the continuum scattering in ordered phases close to a fractionalized phase, because their mean field starting point corresponds to a free spinon gas. The Schwinger boson formalism~\cite{Arovas88,Read91,Chubukov94} offers a natural framework where spinons, coupled through emergent gauge fields, are treated as bosons. This allows for a unified description of both ordered and spin liquid phases. In this picture, long-range magnetic order arises from spontaneous condensation of the Schwinger bosons and magnon modes are obtained as two-spinon bound states induced by fluctuations of the gauge fields~\cite{Ghioldi18}. 
 
 Because Schwinger boson theory is formulated on the lattice, it captures both short- and long-wavelength limits, enabling a comprehensive description of inelastic neutron scattering data. A comparison with recent experiments~\cite{Macdougal20,Ghioldi22}, (which is reproduced in Fig.~\ref{fig:fig1}~(c))  confirms that these calculations can accurately reproduce the magnon dispersion of Ba$_3$CoSb$_2$O$_9$, as well as the structure of the anomalous scattering continuum. 
 
 Following these findings, efforts to find TLHAF quantum spin liquids  ~\cite{Xie23,scheieProximate2024,scheieSpectrum2024} have compared inelastic neutron scattering data from a series of Yb delafossites, AYbSe$_2$ (A being an alkali metal), with tunable $J_2/J_1$ ratios, with state-of-the-art calculations of the dynamical spin structure factor using both DMRG and Schwinger boson calculations (see Fig.~\ref{fig:fig1}~(d)). Access to these state-of-the-art computational methods --- and their agreement with observations --- plays a crucial role in interpreting complicated neutron scattering spectra. Here again, advanced analytic theory and the large-scale quantum many-body simulation work with experiments in an integrated manner.
 
 Once such study~\cite{scheieSpectrum2024} recently reported that NaYbSe$_2$ exhibits no detectable phase transition down to the base temperature of the experiment. Unfortunately, it would be premature to draw definitive conclusions about the ground state, as the mobility of Na ions can introduce significant structural and magnetic disorder~\cite{cairns2024}, potentially obscuring intrinsic quantum behaviour. 

However, if a robust realization of a spin liquid is confirmed, there remains the challenge of identifying its character among many competing theoretical scenarios. The observation of a low-energy spin gap in the inelastic neutron scattering spectrum would rule out a gapless $\mathbb{U}(1)$ Dirac quantum spin liquid~\cite{Kaneko14,Iqbal16}. Nonetheless, several distinct gapped phases have been proposed for the triangular lattice: a resonating valence bond (gapped $\mathbb{Z}_2$) spin liquid\cite{Sachdev92,Wang06,Shackleton25}, a valence bond crystal ~\cite{Miksch21,Seifert2024}, and a chiral quantum spin liquid~\cite{Szasz20}.

Both the chiral quantum spin liquid and the valence bond crystal break discrete symmetries—time-reversal and lattice symmetries, respectively and therefore must exhibit finite-temperature phase transitions to restore these symmetries at higher temperature. In contrast, the $\mathbb{Z}_2$ spin liquid preserves all symmetries and does not undergo such a transition. The characteristic of the $\mathbb{U}(1)$ Dirac quantum spin liquid, on the other hand, is a gapless continuum at $K$ and $M$ points of the Brillouin zone.

This rational approach --- combining high-resolution inelastic neutron scattering measurements with advanced numerical and analytical techniques --- is proving highly effective in guiding the search for extreme quantum states of matter, such as spin liquids. In particular, it provides direct evidence for the proximity to, or presence of, fractionalization, which is essential for identifying and characterizing these phases.

As well as providing support for experimental claims of spin liquid behaviour, numerical calculations are also increasingly playing a role in ruling out spin liquid candidates. For example, the combination of ground-state phase diagram calculations using DMRG~\cite{Zhu18} and ultralow-temperature a.c. susceptibility measurements~\cite{maSpinGlass2018} revealed that YbZnGaO$_4$, previously proposed as a triangular-lattice spin liquid candidate, is in fact in a spin glass state.

\begin{figure*}[htp!]
\includegraphics[width=\textwidth]{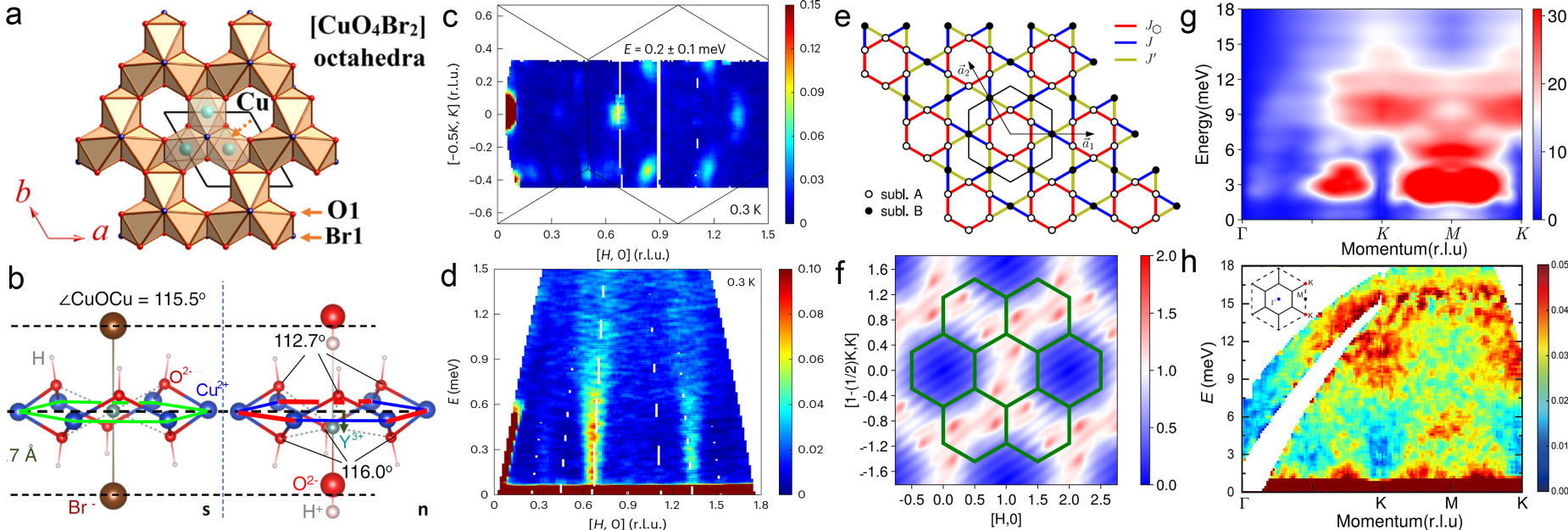}
\caption{\textbf{Kagome lattice antiferromagnet YCu$_3$-Br.} \textbf{a}, The structure of a kagome layer in YCu$_3$-Br, consisting of [CuO$_4$Br$_2$] octahedra \cite{ChenXH20}. \textbf{b}, Optimized local structures for supercells containing symmetric (s) and nonsymmetric (n) environments obtained from density functional theory calculations \cite{LiuJ22}. Y$^{3+}$ ions reside at hexagonal centers but occupy two nonequivalent sites, within and displaced the kagome planes. This positional distinction arises because some of the Br$^-$ ions connecting Y$^{3+}$ along the out-of-plane $c$-axis are substituted by (OH)$^-$, which is polar and pulls the connected Y$^{3+}$ ion out of the kagome plane. The different valences and ion sizes between Y$^{3+}$ and Cu$^{2+}$ help eliminate weakly-correlated magnetic impurities often observed in herbertsmithite and related compounds due to partial substitution of Cu$^{2+}$ ions on nonmagnetic ions~\cite{NormanMR16}. \textbf{c and d}, Intensity contour plot of the inelastic neutron scattering results in the [$H$,$K$] zone at 0.2 meV and as a function of energy, $E$, and momentum, $Q$, along the [$H$,0] direction, respectively \cite{ZengZ24}. \textbf{e}, Schematic illustration of the 3$J$ model used for classical spin wave calculations\cite{HeringM22}, which introduces three different nearest neighbour coupling constants between sites. Note that the unit cell of this model is three times larger than that of the kagome lattice, as represented by the black hexagon. \textbf{f}, The static spin structure factor $S(Q)$ obtained from density matrix renormalization group calculations as described in the main text. The green hexagons depict the kagome Brillouin zone. \textbf{h and g}, Spin excitations from density matrix renormalization group calculations and inelastic neutron scattering measurements, respectively \cite{hanSpin2024}. The inset in \textbf{h} illustrates high-symmetry points in the kagome Brilluoin zone.}  
\label{fig:Dirac}
\end{figure*}

\section*{\noindent{\textbf{Kagome lattice Heisenberg antiferromagnets}}} 

One of the most challenging issues in frustrated quantum magnetism is understanding the kagome lattice Heisenberg antiferromagnet. This model has been an long-term candidate quantum spin liquid~\cite{NormanMR16}. Despite extensive theoretical modeling, large-scale simulations, and experimental investigations, however, no consensus has emerged on whether any specific material definitively realizes a quantum spin liquid state, and if so which kind, even for the simplest model with only nearest-neighbour exchange interactions. 


This debate extends to key kagome quantum spin liquid candidates, such as herbertsmithite and Cu$_3$Zn(OH)$_6$FBr \cite{hanFractionalized2012,FengZL17,khuntia2020}. However, progress is hindered by intrinsic magnetic impurities from a small percentage of Cu$^{2+}$ ions on nonmagnetic sites. It is complicated to disentangle impurity effects from the intrinsic low-energy excitations that would reflect spin liquid behaviour~\cite{NilsenGJ13,WeiY21}. 

Beyond impurity effects, direct comparisons between theory and experiment remain difficult, as real materials typically include additional, non-Heisenberg interactions beyond the theoretical ideal modes; and their high-energy spin excitations are rarely observed due to the weak signal of the broad excitations, adding complexity to theoretical calculations.

Following advances in material synthesis and experimental methods, possible evidence for gapless, linearly-dispersing Dirac spinons has been identified in YCu$_3$(OH)$_6$Br$_2$[Br$_{1-x}$(OH)$_x$] (YCu$_3$-Br)~\cite{ZengZ22,ZengZ24,JeonS24}. The YCu$_3$-Br structure consists of ideal kagome planes formed by Cu$^{2+}$ ions, each of which is surrounded by four equatorial O$^-$ and two apical Br$^-$ (B1) anions, forming a [CuO$_4$Br$_2$] octahedron, shown in Fig.~\ref{fig:Dirac}\textbf{a} ~\cite{ChenXH20}. 
These developments offer a fresh direction to address long-standing questions in the study of kagome Heisenberg antiferromagnets.


Many studies of YCu$_3$-Br have demonstrated the absence of magnetic ordering down to 50 mK\cite{ChenXH20,ZengZ22}. Evidence for Dirac spinons first emerged in specific heat results, which display a $T^2$ dependence at zero field and an additional linear term at high magnetic fields~\cite{ZengZ22}. These behaviours align with mean-field predictions for a U(1) Dirac quantum spin liquid~\cite{RanY07}. Raman spectra further reveal broad magnetic excitations at high energies, attributed to one-pair and two-pairs of spinon-antispinon excitations~\cite{JeonS24}. The sublinear power-law dependence of the Raman susceptibility on energy also suggests Dirac spinons at high energies. 

More compelling evidence comes from inelastic neutron scattering~\cite{ZengZ24}, which shows a conical spin continuum at low energies (Figs.~\ref{fig:Dirac}\textbf{c} and ~\ref{fig:Dirac}\textbf{d}). The width of these spin excitations demonstrates a precise linear dependence on energy, consistent with a Dirac spinon cone centered at or near zero energy. Notably, the estimated spinon velocity accurately reproduces the low-temperature specific heat without further parameter adjustments. 

These features differ markedly from semiclassical spin-wave calculations, where introducing strong randomness to generate a continuum-like spectra broadens low-energy spin excitations and reduces their central intensity~\cite{ChatterjeeD23}. Crucially, a damped spin-wave model cannot account for the substantial intensity observed in neutron scattering spectra at the two-dimensional kagome Brillouin zone point (1/3,0) due to the structure factor constraints \cite{hanSpin2024}.

Although YCu$_3$-Br possesses an ideal Cu$^{2+}$ kagome lattice, the location of its spin excitations in momentum space, as measured by neutron scattering, cannot be reproduced by conventional Heisenberg models. This discrepancy arises from the distorted local structure (Fig.~\ref{fig:Dirac}\textbf{b}), which generates two distinct hexagonal structures. Consequently, three different nearest exchange interactions emerge. 

The observed quantum-spin-liquid-like behaviour emerges when the proportion of one of these structures reaches around around 2/3 \cite{XuA24}. Furthermore, the sharp low-energy spin excitations of YCu$_3$-Br demonstrate that the two local structures are  not randomly distributed.

In a similar material Y$_3$Cu$_9$(OH)$_{19}$Cl$_8$ (Y$_3$Cu$_9$-Cl), the two hexagonal structures arrange in an ordered pattern, forming a kagome superlattice with three times enlarged unit cell area ~\cite{ChatterjeeD23}. Both YCu$_3$-Br and Y$_3$Cu$_9$-Cl exhibit identical low-energy magnetic excitation wavevectors despite the latter forming an antiferromagnetically ordered phase~\cite{ChatterjeeD23}, indicating they necessitate comparable theoretical descriptions. 


\emph{Ab initio} density functional theory calculations have identified three distinct nearest-neighbour superexchange interactions in Y$_3$Cu$_9$-Cl~\cite{HeringM22}. A classical spin-wave analysis of the kagome Heisenberg model incorporating these couplings (Fig.~\ref{fig:Dirac}\textbf{e}, known as the 3$J$ model) captures the antiferromagnetic order observed in Y$_3$Cu$_9$-Cl~\cite{HeringM22,ChatterjeeD23}. Yet this description departs from the ideal kagome Heisenberg limit, as the arrangement of the distortions breaks translational symmetry. 

This raises several theoretical and computational challenges. Given the similarities between YCu$_3$-Br and Y$_3$Cu$_9$-Cl, there is hope that, starting from a common theoretical model of the two compounds  \emph{ab initio} methods, density matrix renormalization group, quantum Monte Carlo and field-theoretical approaches such as Schwinger boson technique that account for strong quantum fluctuations may be able to identify a possible transition from antiferromagnetic order to a quantum spin liquid. This should make it possible to track the evolution of fractionalized excitations even when the ground state is not a spin liquid, providing essential guidance for experiments. 

The freedom to explore theoretical parameter space might also make it possible to search for a quantum spin liquid as a stable intermediate phase on the path between the Y$_3$Cu$_9$-Cl structure with hexagonal distortions, towards the theoretical ideal homogeneous kagome Heisenberg antiferromagnet~\cite{yanSpin2011,DepenbrockS12,RanY07,liaoGapless2017}. At present, fundamental questions remain open and call for systematic exploration.

An initial integrated numerical study of the 3$J$ model---combining classical Landau–Lifshitz dynamics with DMRG time evolution---has already succeeded in reproducing the static structure factor (Fig.~\ref{fig:Dirac}\textbf{f}) and the high-energy spectral features (Figs.~\ref{fig:Dirac}\textbf{g},\,\textbf{h}) observed experimentally~\cite{hanSpin2024}. The influence of the Dzyaloshinskii–Moriya interaction on the low-energy spin excitations has likewise been examined using DMRG, and the results capture the enhanced in-plane magnetic fluctuations revealed by recent polarized neutron-scattering measurements~\cite{hanSpin2024}. 

This agreement reinforces the validity of the 3$J$ model and highlights the need for more advanced computational approaches to resolve the remaining low-energy excitation spectrum. 
Intriguingly, similar spin liquid behaviour in specific heat at zero and finite magnetic field has also been reported in LuCu$_3$(OH)$_6$Br$_2$[Br$_{0.33}$(OH)$_{0.67}$] (LuCu$_3$-Br)~\cite{LiZ25}, which shares the same nuclear structure as Y$_3$Cu$_9$-Cl. On the other hand, LuCu$_3$-Br does not have the randomness of the two local hexagonal structures present in YCu$_3$-Br. This demonstrates that randomness or disorder is not a necessary ingredient for realizing quantum spin liquid physics within this family of models.\\


\noindent{\textbf{\large Discussion}}

The three examples highlighted here mark recent milestones in quantum magnetism and underscore a common theme: progress emerges from a balanced integration of advanced numerical simulations, field-theoretical analysis and materials synthesis and characterization. This collaborative framework moves beyond the outdated view that one pillar---such as predictions from a toy model---drives discovery whereas the others merely validate it. Instead, it affirms their equal importance in advancing our understanding of quantum materials.


TRG and DMRG techniques, augmented with state-of-the-art ab initio methods now allow simulations of realistic spin model Hamiltonians on cylindrical geometries. This capability enables access to thermodynamic properties, such as magnetic susceptibility and specific heat, for quantitative comparisons with experimental data. Furthermore, these comparisons with experimental data can be refined into an iterative scheme, where optimized model parameters are determined through automated, machine-learning-based software, paving the way for more precise and efficient model fitting. 

Most tensor network calculations of spectra and thermodynamic properties are still performed using a quasi-one-dimensional cylindrical geometry in DMRG, as this is the only setting in which the scaling of the needed computational resources with system size can still be controlled. However, the limitations of this approach, and the need to perform true two-dimensional tensor-network renormalization group calculations, have become more and more obvious, if one would like to have good momentum resolution in the two-dimensional Brillouin zone. Achieving this with acceptable accuracy and an affordable computation burden, would enable a much more detailed comparison between experiment and model calculations, substantially strengthening the integrated approach.


With access to optimized model parameters, QMC simulations can then compute the partition function of fully optimized two-dimensional lattice model Hamiltonians and access all dynamical and thermodynamic correlation functions with the lower computational complexity and access to both ground state and finite temperature that is difficult to achieve using  tensor network methods. Until recently, QMC methods were limited by their reliance on imaginary-time data in the path-integral framework, which cannot be directly compared to real-time or frequency dependent experimental data. However, this limitation has been effectively addressed by the stochastic analytic continuation technique~\cite{shaoNearly2017}, which transforms a laborious search for the optimized spectrum in an exponentially large space into a structured, importance-sampling Monte Carlo process. This procedure allows QMC to yield real-frequency dynamic spectra that can be directly compared with inelastic neutron scattering and NMR experimental results~\cite{liKosterlitz2020,huEvidence2020,liaoPhase2021}. 

In the future, better lattice models are needed so that QMC simulations can capture the complex physics of real materials, including multiple interactions and geometric frustration. Most direct implementations of these properties with QMC will encounter the minus sign problem, or very long autocorrelation times in the update scheme, making calculations intractable. One needs to find effective models that capture the essential physics of the material and yet remain numerically managable using QMC. 

One recent advance in this direction is the  ability to directly simulate emergent gauge fields coupled to fractionalized excitations on the lattice~\cite{xu2019monte,chenEmergent2025}. Such simulations can provide valuable insight into quantum spin–liquid physics across a range of temperatures and magnetic fields, offering a powerful framework for guiding and interpreting experiments.


The success of the Schwinger boson formalism in interpreting the inelastic neutron scattering spectra of Ba$_3$CoSb$_2$O$_9$~\cite{Macdougal20,Ghioldi22} demonstrates the continued power of analytical methods. Although parton theories can capture key features of spectroscopic measurements, going beyond mean-field theory is crucial to accurately describe both the collective modes and the broad scattering continuum on the magnetically ordered side.

More advanced Schwinger boson calculations already incorporate low-order contributions from gauge-field fluctuations~\cite{Ghioldi18,Ghioldi22}. This extension has led to significantly improved agreement with experimental data, indicating that the systematic inclusion of higher-order fluctuation contributions is a promising direction for further development.
More generally, within large-$N$ parton frameworks, incorporating fluctuations beyond the mean-field approximation is essential for describing collective modes with integer spin and, more broadly, for achieving an accurate characterization of dynamical correlations~\cite{willsher2025}.

One of the biggest challenges facing the field of quantum magnetism, which we have not yet discussed, is the ongoing search for Kitaev quantum spin liquids. Following Alexei Kitaev’s exact solution of the honeycomb model~\cite{Kitaev2006}, substantial effort has gone into identifying materials that might realize this physics~\cite{Jackeli2009}, with candidates including various iridates and ruthenates~\cite{Trebst2022}. 

Despite extensive research into these candidate materials, the details of the appropriate spin Hamiltonians to capture their behaviour remain elusive~\cite{moller2025,matsudaKitaev2025}. This uncertainty stems from multiple factors. For example, the theoretically ideal Kitaev interaction is often accompanied by additional anisotropic exchange terms. The magnetism in the candidate materials arises from $4d$ and $5d$ orbital electrons, whose spatial extent introduces further-neighbour couplings. Furthermore, there is evidence that accurate modeling of the electronic structure may require more careful treatment of the hybridization between ligand and metal orbitals than has been performed so far. 

Given the large number of symmetry-allowed interaction terms, a reliable extraction of the effective spin model requires a multitarget approach that synthesizes different experimental observables with predictive many-body simulations~\cite{moller2025,matsudaKitaev2025}. As it stands, the Kitaev model is a beautiful but abstract example of emergent gauge theories in spin systems. Absent a comprehensive strategy anchored in both experiment and computation it may remain as speculative as the original lattice gauge theories of high-energy physics~\cite{Kogut1979}. 

This is where the more integrated approach we advocate in this article could improve the situation. Ab initio methods and TRG can be iteratively compared with thermodynamic measurements to determine the precise model parameters. With access to an effecitve model, more accurate techniques such as QMC can be used to make quantative comparisons with inelastic neutron scattering results. In this way, the successes achieved in the three examples could be replicated to advance the study of Kitaev physics.

We encourage the community to adopt a balanced perspective in which advanced numerical simulations, cutting-edge field theory and innovative materials synthesis and characterization operate in concert rather than in hierarchy. Such integration is essential for uncovering deeper insights into quantum systems and, ultimately, for achieving the long-sought experimental breakthroughs needed to establish quantum spin liquids conclusively.\\

\end{CJK*}
\bibliography{bibtext}

\vspace{0.5cm}
\noindent{\textbf{\large Acknowledgments}}\\
ZYM thanks Chengkang Zhou for helping with making the figures and acknowledge the support from the
Research Grants Council (RGC) of Hong Kong (Project Nos. AoE/P701/20, 17309822, C7037-22GF, 17302223, 17301924, 17301725), the ANR/RGC Joint Research Scheme sponsored by RGC of Hong Kong and French
National Research Agency (Project No. A\_HKU703/22) and the State Key Laboratory of Optical Quantum Materials at HKU. 
We thank HPC2021 system under the Information Technology Services and the Blackbody HPC system at the Department of Physics, University of Hong Kong, as well as the Beijng PARATERA
Tech CO.,Ltd. (URL: https://cloud.paratera.com) for providing HPC resources that have contributed to the research results reported within this paper. CDB acknowledges support by the U.S. Department of Energy, Office of Science, National Quantum Information Science Research Centers, Quantum Science Center.  SL acknowledge the support from the National Key Research and Development Program of China (Grants No. 2022YFA1403400, No. 2021YFA1400400), the Chinese Academy of Sciences (Grants No. XDB33000000, No. GJTD-2020-01). ZYM and CDB are grateful to the Pollica Physics Centre for hospitality during the stimulating workshop “Exotic quantum matter: from quantum spin liquids to novel field theories” (2024). \\












\end{document}